\let\FONTS=C
\catcode`\{=1 
\catcode`\}=2 
\catcode`\#=6 
\let\x=\input
\def\y#1 {\let\input\x \let\x=\undefined \input lfonts.new}
\def\input#1 {\let\input=\y \let\y=\undefinded}
\x lplain

\input clmult01.mac
\makeatletter
\input amsfont.sty
\input cropmark.sty
\nopagecropped
\makeatother

\documentstyle[amssymb]{clmult01}
\input{epsf}

 \newcommand{\bm}[1]{\mbox{\boldmath $#1$}}

 \newcommand{\zr}[1]{\mbox{\hspace*{#1em}}}

 \newcommand{\EE}{{\Bbb E}}
 
 \newcommand{\RR}{{\Bbb R}}
 \newcommand{\ZZ}{{\Bbb Z}}

 \newcommand{\Id}{\mbox{\rm 1\zr{-0.62}{\small 1}}}

\begin{document}

\title{A Guide to Mathematical Quasicrystals}
\author{{\sc Michael Baake}}
\institute{Institut f\"ur Theoretische Physik, Universit\"at T\"ubingen, \\
           Auf der Morgenstelle 14, D-72076 T\"ubingen, Germany} 
\maketitle


\section{Introduction}

The discovery of alloys with long-range orientational order and
sharp diffraction images of non-crystallographic symmetry \cite{Shecht,INF}
has initiated an intensive investigation of the possible structures and physical
properties of such systems. Although there were various precursors,
both theoretically and experimentally \cite{Suck}, it was this renewed and
amplified interest that established a new branch of solid state physics, and also
of discrete geometry. It is usually called the theory of quasicrystals, even though
it also covers ordered structures more general than those with pure Bragg diffraction
spectrum.

It is now rather common to think of the regime between
crystallographic and amorphous systems as an interesting area with a hierarchy 
of ordered states. This was not so some fifteen years ago, and it is the purpose
of this contribution, and of the book as a whole, to introduce some of the ideas
and methods that are needed to handle this new zoo. In particular, I will summarize
some mathematical and conceptual issues connected with it, with special emphasis
on proper equivalence concepts. This is more important than it might appear at first
sight, because non-periodic order shows both new features and new hazards -- and
it is worthless to talk about a property of one specific structure if it is lost
for others that are locally indistinguishable.

To develop some of these ideas, one has to start with a valid idealization of
the physical structures one has in mind. Since we are interested in solids
of some relevant size here, it is reasonable to replace their atomic arrangements 
by suitable {\em infinite\/} point sets. These should be {\em uniformly discrete\/} 
(i.e., there should be a uniform minimal distance
between the points) and, usually, they should be 
{\em relatively dense\/} (i.e., there is a maximal hole). 
Sets with this property are called {\em Delone sets\/} and are widely
used for this purpose. \index{Delone set}

In analogy to ordinary crystallography, many people prefer to think in terms of
cells or {\em tiles} \cite{NB}. 
Here, one may start from a (usually finite) number of proto-tiles
that fit together   
to tile space without gaps or overlaps. 
If we now decorate the tiles by finitely many points 
(giving the atomic positions, say), we return to a Delone set. 
Vice versa, given a Delone set $\Lambda$, we can perform the Voronoi
construction that attaches to each point $x\in\Lambda$ the region of all points
of ambient space that are closer to $x$ than to any other point of $\Lambda$. 
This way, we come back to a tiling (whose dual, the so-called Delone tiling \cite{Sch}
is an even better candidate). Under an additional, but rather mild, condition, 
namely that $\Lambda-\Lambda$ is discrete\footnote{A point set $S$ is {\em discrete},
if each point $x\in S$ can be surrounded by a non-empty open ball that does not contain 
any point from $S$ except $x$. Note that uniform discreteness is a lot stronger
than mere discreteness!}, 
it will actually show only finitely many different tiles, and in this sense the 
two concepts are equivalent \cite{Jeff}.

In what follows, in line with other expositions \cite{KG,KS,Dun},
we shall usually illustrate the concepts with examples from
the class of tilings made from finitely many proto-tiles, 
but all concepts will be formulated in a way that allows
to switch between such tilings and Delone sets.
In particular, the central equivalence concepts will be described in that way,
and it will actually make this very connection between tilings and point
sets more precise.

As mentioned before, it is the aim of this contribution to summarize several 
rather central concepts needed for the description of quasicrystals. 
Since it is impossible to give a self-contained exposition of the
present body of knowledge in this survey, the following text is often a pointer 
in nature, and further details have to be taken from the references given. 
Also, even though
some definitions and results are high-lighted, the exposition is rather informal
and contains no proofs, though we try to sketch the ideas behind them,
whenever that is possible in fairly non-technical terms.

The article is organized as follows. After an introduction to
non-crystallo\-graphic symmetries, we will recall some concepts from
{\em diffraction theory\/}, followed by a Section on quasiperiodicity 
and the {\em projection method}, 
which is vital for the description of perfect quasicrystals. 
We then explain the issue of {\em minimal embedding}, illustrated with 
the most important examples. Then, as a first equivalence concept, 
{\em local indistinguishability} (also called {\em local isomorphism})
in introduced.
This allows for the proper definition of (generalized) symmetry, but also calls
for an effective control of the corresponding equivalence classes, which is
then provided by the {\em torus parametrization}.

This is followed by the introduction of the important equivalence
concept of {\em mutual local derivability}, or {\em local equivalence}. 
This paves the way for a generalization of translation and space group
concepts from ``ordinary'' crystallography, without using Fourier transforms.
Also, this setting enables a simple and unified setting for 
{\em inflation/deflation symmetries} and {\em perfect matching rules} as invariants of
local equivalence classes. After some speculations on a possible classification
scheme (which is presently still pretty incomplete), a brief introduction to
the alternative approach via so-called ``random tilings'' is given, which is, in
a way, the stochastic counterpart of the theory of perfect quasiperiodic tilings.

\section{Non-crystallographic symmetries}

To understand the impact of the discovery of quasicrystals, we first have to
know what non-crystallographic symmetries are (to be described in this Section) 
and then why their appearance in the diffraction images of solids is, at least at
first sight, astonishing (to be addressed in the next Section).
\index{non-crystallographic symmetries}

\begin{figure}[ht]
\vspace*{3mm}
\centerline{ \epsfysize=80mm 
               \epsfbox{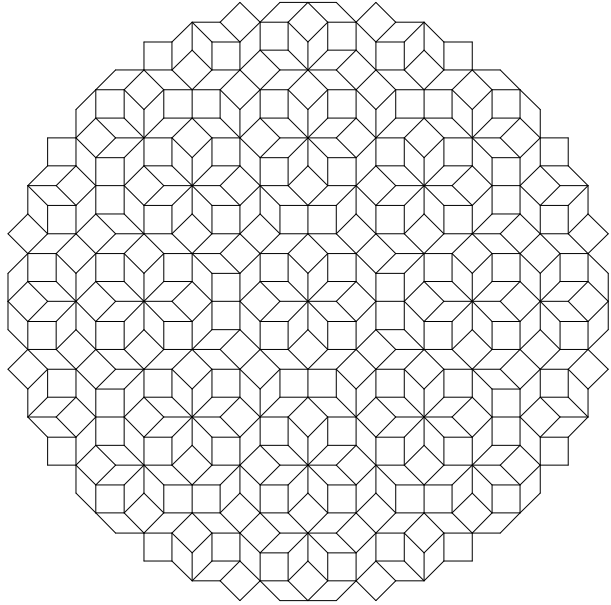}}
\vspace*{5mm}
\caption{Central patch of an exactly eightfold symmetric Ammann-Beenker tiling.}
\label{ab}
\end{figure} 

{}For a start, we need some crystallographic concepts, in particular that of
a lattice in $n$-dimensional ($n$-D) Euclidean space. 
A set $\Gamma\subset\RR^n$ is called a {\em lattice \/} if it is a 
discrete subgroup of $\RR^n$ such that the factor group
$\RR^n/\Gamma$ is compact. This is equivalent to saying that there is a
set of linearly independent vectors, $\bm{b}_1, \ldots, \bm{b}_n$, called the 
{\em basis} of the lattice, such that  \index{lattice}
\begin{equation}
     \Gamma = \ZZ\bm{b}_1 \oplus \cdots \oplus \ZZ\bm{b}_n \, ,
\end{equation}
i.e., $\Gamma$ consists of all integer linear combinations of the basis vectors.
So, $\ZZ$ is a lattice in one dimension, and $\ZZ^2$ is one in the plane, but $\ZZ$ 
is {\em not\/} a lattice in two or more dimensions because its basis only spans a
1D ambient space.

Next, we call a set $S$ {\em periodic}, if $S+\bm{t} = S$ for some $\bm{t}\neq 0$.
Such a $\bm{t}$ is a {\em period} of $S$. $S$ is called {\em crystallographic} if 
its periods form a lattice, i.e.\ if its periods span ambient space, compare
\cite{petermatch}. \index{crystallographic} \index{periodic}
If we now ask for the possible rotation symmetries of a crystallographic point set, 
we hit an obstruction, usually called the {\em crystallographic restriction\/}
\cite{Schwarz}.
\begin{theorem}
   Let $S$ be a crystallographic point set in $\RR^n$, and $R$ an orthogonal 
   transformation that maps $S$ onto itself. 
   Then, $R$ is of finite order, i.e.\ $R^k = \Id$ for some $k$. 
   In particular, we have $k\in\{1,2\}$ for $n=1$, and
   $k\in\{1,2,3,4,6\}$ both for $n=2$ and $n=3$. 
   In general, the characteristic polynomial of $R$ has integer coefficients only.
\end{theorem} \index{crystallographic restriction}
The reason is that the invariance of $S$ implies the
invariance of its lattice of periods. The discreteness of the lattice forbids
$R$ to be of infinite order. Next, $R$ must map the basis vectors onto
integer linear combinations of them, so it must be similar to an integer matrix,
from which the last statement follows. The case $n=1$ is trivial,
while $n=2$ and $n=3$ results from considering the traces of orthogonal matrices
in their standard form: if $\varphi$ is the rotation angle
(around an axis in 3D), then $2 \cos(\varphi)$ must be an integer.

\begin{figure}[ht]
\vspace*{3mm}
\centerline{ \epsfysize=85mm 
               \epsfbox{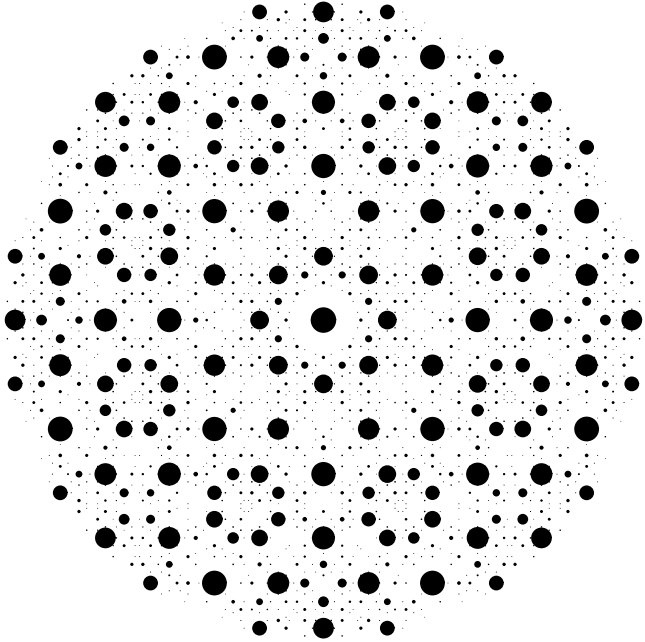}}
\vspace*{5mm}
\caption{Diffraction image of the Ammann-Beenker tiling of Figure~\protect{\ref{ab}}. 
The area of each disc is proportional to the intensity of the corresponding peak, 
the cut-off is at 0.1 $\%$ of the central intensity.} \label{ab-fou}
\end{figure}

\index{Ammann-Beenker tiling}
As a consequence, the so-called Ammann-Beenker tiling of Figure~\ref{ab}, showing 
exact eightfold symmetry even in the infinite area limit where it covers the plane, 
cannot be crystallographic \cite{Ammann,GS,BJ}. 
What is more, it cannot possess {\em any\/} period,
because the symmetry would immediately complete any single period
to a basis of a lattice.
So, this is not a model of a 2D crystal, neither is any atomic decoration of its 
two cells, a square and a rhomb. The simplest such decoration, which is typical
and exhaustive in a sense we will understand shortly, consists of points 
on all vertices of the entire tiling. If we take a diffraction image of this,
with scatterers of equal strength on all points, we nevertheless obtain an image
that resembles that of crystals pretty closely, see Figure~\ref{ab-fou}.
Let us thus turn our attention to diffraction for a while.

\section{Diffraction}

To simplify things, we will only talk about kinematic diffraction, i.e.\
diffraction that can be understood in terms of single scattering in the
Fraunhofer picture. This is quite appropriate for X-ray and neutron diffraction,
but not for electron diffraction where multiple scattering is essential, see
\cite{Cowley} for details. Kinematic diffraction from a structure, in turn, 
is closely related to the Fourier transform of the corresponding potential
in the sense that the observed intensities of sharp spots (Bragg peaks) are 
proportional to the absolute squares of the Fourier amplitudes, see \cite{Cowley}
for a more detailed discussion and justification of this point of view. 
\index{diffraction} \index{Bragg peaks}

In our idealized world, with atoms etc.\ replaced
by point scatterers (Dirac distributions) on a set $\Lambda$, we thus consider
the so-called {\em Dirac comb\/} of $\Lambda$, \index{Dirac comb}
\begin{equation}
    \omega \; = \;  \omega^{}_{\Lambda} \; := \; \sum_{x\in\Lambda} \delta_x \, ,
\end{equation}
where $\delta_x$ is Dirac's distribution (or measure) at point $x$, i.e.\
$(\delta_x,\psi) = \psi(x)$ for all test functions $\psi(x)$. 
In a second step, one defines a so-called {\em autocorrelation\/} or
{\em Patterson function\/} \cite{Cowley} for this, \index{autocorrelation} 
\index{Patterson function}
\begin{equation}
    \gamma^{}_{\omega} \; := \; \lim_{r\rightarrow\infty}
      \frac{1}{{\rm vol}(B_r(0))}
      \sum_{x,y\in\Lambda\cap B_r(0)} \delta_{x-y} \, ,
\end{equation}
where $B_r(0)$ is the (solid) ball of radius $r$ around $0$, and the limit
is assumed to exist. Furthermore, we will tacitly assume that this limit stays
the same if we replace the ball by any other convex region centred around 
the origin (with $r$ the radius of the maximal inscribed ball) -- an actually rather
non-trivial feature to establish. The autocorrelation is a distribution of
the form
\begin{equation}
      \gamma^{}_{\omega} \; = \; \sum_{z\in\Delta} \nu(z) \delta_z
\end{equation}
where $\Delta = \Lambda-\Lambda$ and $\nu(z)$ is the density of points
$x\in\Lambda$ such that also $x+z\in\Lambda$. In particular, the coefficient
of $\delta_0$ is $\nu(0)=d={\rm dens}(\Lambda)$.

The observed intensity pattern is now given by the Fourier transform of this
autocorrelation, denoted by $\hat{\gamma}^{}_{\omega}$. Here, the following 
convention for the Fourier transforms of functions is used:
\index{Fourier transform}
\begin{equation}
   \hat{f}(k)   \; = \; \int_{\RR^n} e^{-2\pi ikx} f(x) dx \; , \quad
   \check{g}(x) \; = \; \int_{\RR^n} e^{ 2\pi ikx} g(k) dk \, .
\end{equation}
With this convention, one has $\hat{\check{g}}=g$ and $\check{\hat{f}\;}\!\!=f$.
Also, there is no need to distinguish between dual and reciprocal lattice
(the factor $2\pi$ is absorbed into the argument of the exponential function)
and the convolution theorem takes the nice form 
$\widehat{f*g} = \hat{f}\cdot\hat{g}$ where
\begin{equation}
    (f*g)(x) \; = \; \int_{\RR^n} f(x-y) g(y) dy \, .
\end{equation}
Finally, the Fourier transform of a tempered distribution \cite{Rudin}, $T$, is defined 
by $(\hat{T},\psi):=(T,\hat{\psi})$, as usual. This way, the coefficient of
$\delta_0$ in $\hat{\gamma}^{}_{\omega}$ is $d^2$.

Let us take a closer look at the diffraction from a lattice $\Gamma$.
Here, $\Delta=\Gamma$ and, obviously, the autocorrelation coefficients
are $\nu(z)\equiv d = {\rm dens}(\Gamma)$. So
\begin{equation}
     \gamma^{}_{\omega} \; = \; d \, \sum_{x\in\Gamma} \delta_x \, .
\end{equation}
It is now a direct consequence of the so-called Poisson summation formula that 
the diffraction image consists of Bragg peaks (or Dirac peaks) only.
They are supported by the dual (or reciprocal) of the lattice of periods, and hence
distributed in a discrete fashion. Poisson's summation formula reads
\begin{equation} \label{poisson}
      \protect{\widehat{ \sum_{x\in\Gamma} \delta_x }} \; = \;
          d \cdot \sum_{y\in\Gamma^*} \delta_y
\end{equation} \index{Poisson summation formula}
with the {\em dual lattice\/} \index{dual lattice}
$\Gamma^* = \{y \mid x \cdot y \in \ZZ \mbox{ for all } x\in\Gamma\}$.
So, we obtain 
\begin{equation}
     \hat{\gamma}^{}_{\omega} \; = \; d_{}^2 \, \sum_{y\in\Gamma^*} \delta_y \, .
\end{equation}

It is worthwhile to note that Poisson's summation formula (\ref{poisson}) also
provides the Fourier transform of $\omega$ itself, which is well-defined in this case.
Since the autocorrelation is essentially a volume-weighted convolution ($*$) of
$\omega$ with itself, the convolution theorem explains why the coefficients
of $\hat{\gamma}^{}_{\omega}$ are the absolute squares of those of $\hat{\omega}$.
This is usually summarized in a so-called Wiener diagram:
\begin{equation}
 \begin{array}{rcl}
    \omega   & \quad\stackrel{*}{\longrightarrow}\quad & \gamma^{}_{\omega} \\
    & & \\
    \mbox{\tiny FT} \downarrow \, & &   \; \downarrow \mbox{\tiny FT} \\[1ex]
    \hat{\omega} &\quad\stackrel{|.|^2}{\longrightarrow}\quad &\hat{\gamma}^{}_{\omega}
   \end{array} 
\end{equation}
Whenever this situation applies, in the sense that all quantities exist and the diagram
is commutative, things are rather simple. In particular, given the situation of
a lattice, atomic profiles (extended scatterers) or more complicated decorations of a
fundamental domain (multiple atoms per unit cell) can be incorporated by means of
convolutions with $\omega$ and then be processed through the Wiener diagram.
Formally, the same process is then always used (at least for the Bragg part),
but this needs extra justification, and often hard analysis for a proof,
compare \cite{Hof,Boris} and references therein for details.

This procedure seems
perfectly well-defined and robust if applied to the diffraction from {\em finite}
patches or samples, but the problems mentioned are then only shifted to the
question in which sense larger and larger samples show diffraction images that
converge -- again, not at all an easy one.

What other situations give rise to well-defined diffraction? In other words:
which distribution of matter diffracts? The answer to this question is far
from being known. So, we shall only sketch some aspects relevant to the understanding
of the diffraction of quasicrystals and related structures.
There is not much freedom to obtain sharp diffraction spots that are arranged in a
discrete fashion, and this is indeed essentially restricted to diffraction from
crystallographic arrangements (up to deviations of density zero, because the
autocorrelation of a set of positive density is not changed by adding or
removing points of density 0). So, if one restricts attention to such sets,
it is obvious that non-crystallographic symmetries cannot show up. This was the
point of view of ``classical'' crystallography -- and this was challenged by the
discovery of quasicrystals.

\begin{figure}[ht]
\vspace*{3mm}
\centerline{ \epsfysize=80mm \epsfbox{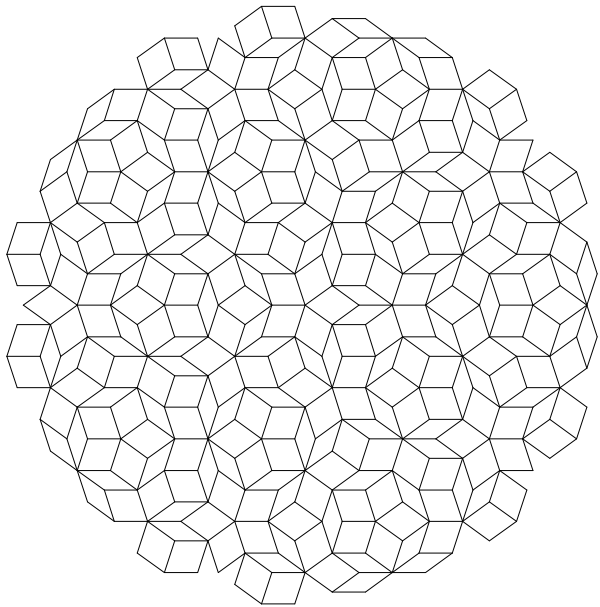}}
\vspace*{5mm}
\caption{Example of a rhombic Penrose tiling.} \label{pen}
\end{figure}

Beyond any doubt, the observed diffraction spots of quasicrystals are sharp
and show non-crystallographic symmetries, such as a fivefold axis or the full
icosahedral group. The solution to the emerging puzzle lies in the answer to the 
question whether the distribution of spots is really discrete\footnote{We only
talk about diffraction from ``single crystals'' here, resp.\ its analogue
for quasicrystals. A somewhat similar discussion appears for the diffraction from
powders, but should be distinguished clearly, because this case has a rather
different explanation, see chapter 16 of \cite{Cowley}.}, i.e.\ whether the
spots are clearly separated from one another. For a given 
resolution, this seems to be the case -- and this is certainly the 
reason why quasicrystals
were discovered (for a given resolution, they show up through a discrete diffraction
pattern). But if one increases the sensitivity (e.g., by doubling the
exposing time of the photo plate), more peaks become visible, and this process
does not come to an end: the set of {\em all} peaks seems to lie dense, and only
those of intensity beyond a given threshold result in a discrete pattern, similar to
the diffraction of a crystal, but with non-crystallographic symmetry.
One example was shown in Figure~\ref{ab}, another important example being
the rhombic Penrose tiling \cite{Penrose,BKSZ,deBr} of Figure~\ref{pen}.
\index{Penrose tiling}

\section{Quasiperiodicity and the projection method}

It is now time to explain how such a strange diffraction behaviour can come 
along. The foundation of it goes back
to the beginning of the century when Harald Bohr, the younger brother of Niels,
developed what is now called the theory of almost periodic functions. Let us
consider the example
\begin{equation}
    f(x) \; = \; \sin(x) + \sin(\tau x)
\end{equation}
where $\tau = (1 + {\scriptstyle \sqrt{5}})/2$ is the famous {\em golden ratio\/}.
This is an irrational number (in fact, as follows from its continued
fraction expansion, the most irrational one), whence $f(x)$ is certainly not periodic. 
Nevertheless, for any given $\varepsilon>0$, there are ``almost-translations''
$t$ such that $|f(x) - f(x+t)|<\varepsilon$, for all $x\in\RR$.
Furthermore, such \index{almost periodic functions}
translations are not rare, but lie relatively dense in $\RR$, i.e.\ there is a
maximal distance between any two consecutive ones. The set of continuous functions 
with this
property is closed under uniform convergence, and can be uniformly approximated
by trigonometric polynomials. This results in a generalization of Fourier series
which is essentially the core of Bohr's work \cite{Bohr}. For a more recent
introduction, with additional material, see \cite{Cor}.
\index{quasiperiodicity}

In these generalized
Fourier series, pairwise incommensurate base frequencies occur (such as 1 and
$\tau$ in the above example). If their number is finite, the corresponding
function is called {\em quasiperiodic\/}. This subclass of functions has the
property that it can be obtained as a section through a periodic function of
more variables, e.g., in our example,
\begin{equation}
        f(x) \; = \; \sin(x) + \sin(y)  |^{}_{y = \tau\, x} 
\end{equation} 
This is also the essential idea to understand the diffractivity of 
quasicrystals, see \cite{kramerneri,duneau,KD} and various articles in \cite{SO}.
\index{projection method} \index{cut and project sets}

\begin{figure}[ht]
\vspace*{8mm}
\centerline{\epsfysize=80mm \epsfbox{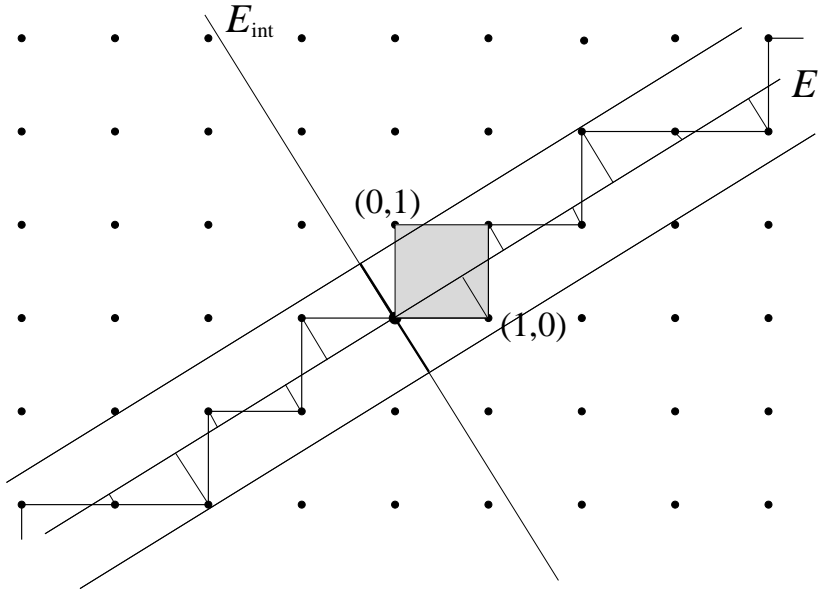}}
\vspace*{5mm}
\caption{Projection method for the Fibonacci chain and torus parametrization
         of its LI-class.} \label{topar}
\end{figure}

Let us therefore construct non-periodic point sets by suitable sections
through a crystallographic structure in higher dimension. As a first step, let
us take a look at the so-called {\em cut and project} method, an example of which
is shown in Figure \ref{topar}. Starting with the square lattice in the plane,
$\ZZ^2$, a line with irrational slope, called $\EE$, is drawn, surrounded by a 
parallel strip of finite width. All lattice points inside the strip are then 
projected to $\EE$. The result is a sequence of points that forms a non-periodic
Delone set (due to the irrationality of the slope -- otherwise it would be 
periodic). If the slope (as in Figure \ref{topar}) is $1/\tau$, and if the width 
of the strip coincides with the projection of a fundamental square to 
the internal direction (which is actually perpendicular here), $\EE_{\rm int}$, 
we obtain what is called the {\em Fibonacci chain}, the most common and best 
studied non-periodic 1D point set. \index{Fibonacci chain}
Note that we have not given a formal definition of a
quasicrystal\footnote{A reasonable working definition of a quasicrystal would include
all discrete patterns which possess an autocorrelation whose Fourier transform
(i.e.\ the diffraction) is either purely discrete, or has at least a
non-trivial discrete part.}, and we will not do so because the present use of
the word is far from being context-free, and a really natural approach is
not yet in sight. Let us add that some authors would prefer not to call the 
Fibonacci chain a real quasicrystal, but rather a modulated crystal.
The reason is topological in nature, compare \cite{Katz}, but since such aspects 
are not important in our present context, we will suppress them. 

This projection scheme, which has an obvious generalization in higher
dimensions, does not seem to be an exact analogue of the section idea mentioned
before, but it is equivalent to it. To see this, take the intersection of the
strip with $\EE_{\rm int}$, which is an interval here. This 
set, $W$, is called {\em window\/} or {\em acceptance domain}, and our Fibonacci
chain $F$ is then given by
\begin{equation}
   F \; = \; \{P(x)\mid x\in\ZZ^2 \;\mbox{ and }\; P_{\rm int}(x)\in W \} \, ,
\end{equation}
where $P$ and $P_{\rm int}$ denote the canonical projections to $\EE$ and
$\EE_{\rm int}$, respectively\footnote{For a more general and systematic
formulation, we refer to \cite{Moody,Martin2,Sch}.}. 
The same set is obtained if we, instead of using
the strip method, take an inverted copy of the window, $-W$, stitch it to each
lattice point of $\ZZ^2$, and modify the rule in saying that we get a point of $F$
whenever our cut line, $\EE$, crosses a copy of this set. It can be considered
as a {\em target} or a kind of {\em atomic hypersurface} which is point-like
in the direction of the ``physical'' space, $\EE$, and extended only in ``internal''
space, $\EE_{\rm int}$. A third method to describe the same object goes under the
name {\em dualization scheme\/} and has the advantage of directly giving cells
rather than point sets, see \cite{ODK,KS,Sch,BKSZ} for details.
\index{dualization method} \index{window} \index{acceptance domain}

In view of the Fourier transform, the version with the atomic hypersurfaces seems
most attractive, because it is closest to the idea of describing a quasiperiodic
arrangement of scatterers as a section through a crystallographic arrangement in
higher dimensions. Consider now a Fibonacci chain, $F$, with point scatterers of
equal strength on all its points, i.e.\ consider the Dirac comb
\begin{equation}
      \omega^{}_F \; = \; \sum_{x\in F} \delta_x \, .
\end{equation}
By a simple (formal) calculation, one finds that the Fourier transform consists
of Dirac peaks on all points $P(k)$ where $k$ is a point of the dual of the
embedding lattice. Its amplitude, $a(P(k))$, is formally given by
\begin{equation} \label{ampli}
     a(P(k)) \; = \; \frac{d}{{\rm vol}(W)}
       \int_{-W} e^{-2\pi i \, k^{}_{\rm int} x^{}_{\rm int}} \, dx^{}_{\rm int} 
\end{equation}
where $x^{}_{\rm int} = P_{\rm int}(x)$ etc.\ and $d={\rm dens}(F)$ 
denotes the density of $F$. The diffraction image is composed of Bragg peaks at 
the points $P(k)$ of intensity $|a(P(k))|^2$.
The derivation of this can be found in many articles,
but has to be taken with a grain of salt: it is purely formal, because the resulting
expression is not a locally summable distribution and hence not a valid representation
of a tempered distribution. That this formal way of calculating 
amplitudes and intensities
is nevertheless correct, was proved much later by Hof, and the
interested reader is referred to \cite{Hof} and references therein.

In what sense does all this resolve the puzzle we started from? If we take
a closer look at Eq.~(\ref{ampli}), we realize that, if $W$ is an interval, the
absolute squares of the amplitudes are of the form 
$\sin(2\pi k^{}_{\rm int})^2/(2\pi k^{}_{\rm int})^2$,
hence bounded by $c/|k^{}_{\rm int}|^2$ with some constant $c$ 
-- and this means that only {\em finitely\/} many peaks per unit volume have an 
intensity beyond a given threshold because $|k^{}_{\rm int}|$ is the distance
of the (dual) lattice point $k$ from the cut space. A cut-off for the intensities 
thus has an effect similar to the projection method itself! This is perhaps one 
of the most important observations in this context: a point set can be diffractive,
and show a clear signature of this, without being crystallographic. If this is the
case, there is then no longer any reason why non-crystallographic symmetries should
not show up. If they do, however, we know immediately that the system cannot be
crystallographic, and we would try to use the idea of a section through a lattice
in higher dimensions to describe the structure.

Let me close this Section with another warning. The success of the projection
method does {\em not} indicate that there is any need for higher-dimensional physics.
It is only a convenient description of a certain class of ordered structures.
Clearly, it is tempting to derive all sorts of generalizations of common properties
and theorems (e.g.\ Bloch's theorem) by employing the embedding scheme and a chain
of formal calculations. Quite frequently, this leads to wrong conclusions, and
extreme care is required. For example, there is no easy analogue of Bloch's theorem.
In fact, its naive generalization fails as badly as possible: the standard
tight-binding model on the Fibonacci chain, in the infinite size limit, has
no bands at all, and the spectrum is neither absolutely continuous nor pure point,
but purely singular continuous! In other words, it is precisely of the form that
was argued impossible for physical structures not too long ago. For more on this,
and on the existence of a Cantor-type gap structure with topological quantum
numbers, see \cite{Suto,Belli,BGJ}. \index{Bloch theorem}
\index{Cantor spectrum}

\section{Minimal embedding and further examples}

In spite of the warnings given, the projection method, if used properly, is an
extremely powerful tool, e.g.\ for practically indexing a diffraction pattern.
But, given a diffractive system with non-crystallographic point symmetry,
what is the right embedding dimension (which equals the indexing dimension)
to start from? Fortunately, the answer is known:
\begin{theorem}
 The description of a planar quasiperiodic point set with $n$-fold symmetry by means
 of the projection method requires a lattice at least of dimension $\phi(n)$,
 where $\phi$ is Euler's totient function\footnote{$\phi(n)$ is the number
 of positive integers less than $n$ which are coprime to $n$.}.
 A 3D quasiperiodic point set with
 icosahedral symmetry requires an embedding lattice at least of dimension 6.
\end{theorem}
The proof of the icosahedral case is based on the representation theory of the
icosahedral group, see e.g.\ \cite{Dun}. 
The statement about the planar symmetries is a direct and rather simple 
consequence of the structure of the so-called
cyclotomic polynomials, see Appendix A of \cite{BJS} for an explicit proof.
\index{minimal embedding}

Let us note that the minimal dimension is usually sufficient (unless one
wants to describe ``modulated'' quasicrystals, where it doubles), and using more
than the minimal number only results in ambiguities of the indexing scheme --
an altogether undesired feature.

Having settled the question for the correct dimension, we need to know
what the ``right'' lattice is. It turns out that the
higher-dimensional analogue of the square and cubic lattices, the hypercubic
lattices, are {\em not\/} sufficient. The most common example where
this becomes apparent is the Penrose tiling of Figure~\ref{pen}. It has
fivefold (actually tenfold) symmetry\footnote{The proper use of the term
symmetry will be explained in the next Section.}, 
and the above Theorem then tells us
that a 4D lattice is the right choice, because $\phi(5)=\phi(10)=4$.
Very often, one finds a description of the Penrose tiling based upon $\ZZ^5$
where one extra dimension has been introduced. This has the disadvantage
mentioned. A simpler choice is the so-called root lattice $A_4$ which
can be seen as the 4D lattice that is obtained by intersecting $\ZZ^5$ with
the 4D hyperplane through the origin, and orthogonal to the space diagonal
$(1,1,1,1,1)$. In general, root lattices provide a very nice class of simple
lattices that is general enough to cover the observed cases \cite{BKJS} in
a maximally symmetric way. For background material on root lattices, and all
sorts of interesting connections to other branches of mathematics, we refer
to the bible, \cite{CS}. \index{root lattices}

Let us briefly mention some other planar examples. The Ammann-Beenker tiling 
of Figure~\ref{ab} shows eightfold symmetry, and requires a 4D lattice
($\phi(8)=4$). The standard choice \cite{BJ} is $\ZZ^4$,
but also the face-centred
lattice in 4D is possible, i.e.\ the root lattice $D_4$. The latter
has the advantage that, with a different choice of the cut space, also
patterns with 12-fold symmetry ($\phi(12)=4$, once more) can be obtained, 
see \cite{BJS} for details. Sometimes, 12-fold symmetry is easier to describe
with another root lattice, namely $A_2\times A_2$.

\begin{figure}[ht]
\vspace*{3mm}
\centerline{ \epsfysize=95mm \epsfbox{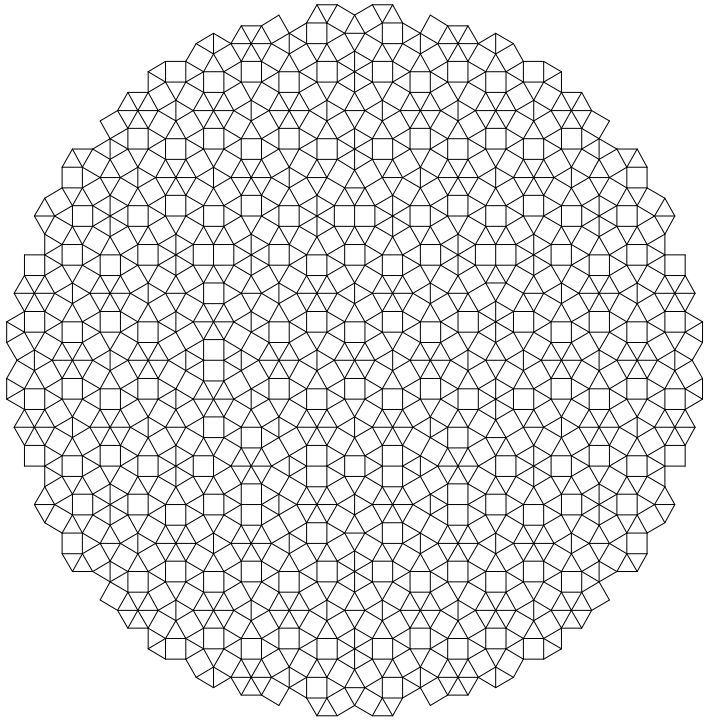}}
\vspace*{5mm}
\caption{Finite patch of a quasiperiodic square-triangle tiling.} \label{stt}
\end{figure}

\begin{figure}[ht]
\vspace*{3mm}
\centerline{ \epsfysize=82mm \epsfbox{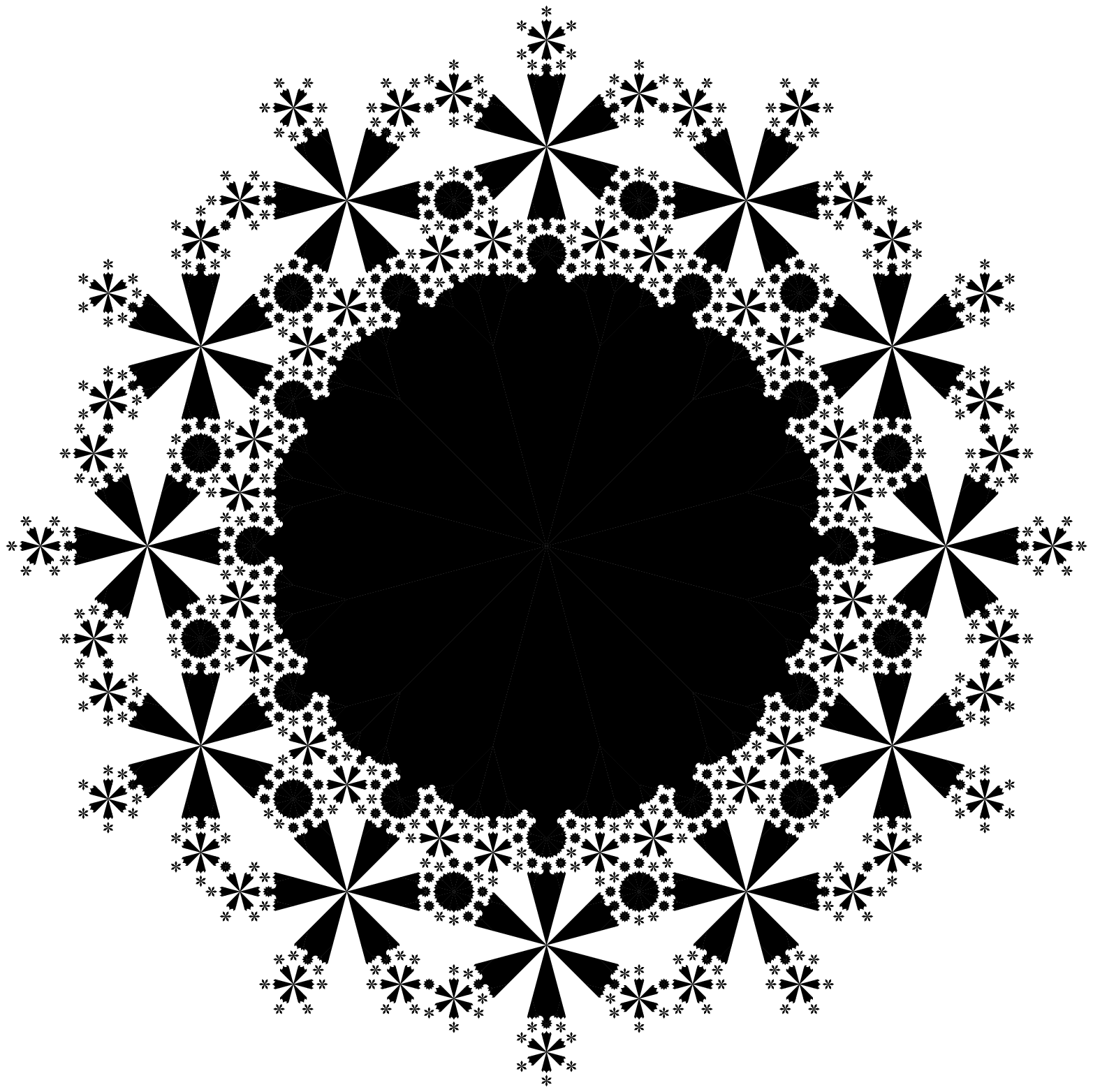}}
\vspace*{5mm}
\caption{Window for the square-triangle tiling of Figure {\protect\ref{stt}}.}
\label{acc}
\end{figure}

Most prominent, in this context, are tilings made from squares and equilateral 
triangles, such as that shown in Figure \ref{stt}. 
It is compatible with 12-fold symmetry (the triangles cover half the area),
and was obtained by the projection method. Its window, however, shows
a more complicated structure: it is a \index{square-triangle tiling}
12-fold symmetric region, compact, the closure of its interior, but
has a fractal boundary, see Figure \ref{acc}.
It is a well-accepted conjecture that all square-triangle tilings with
12-fold symmetry, obtained by projection, require a fractally shaped
window, and, in a certain sense, the one of Figure~\ref{stt} is an example
with ``maximally smooth'' window boundary \cite{BKS}: almost everywhere, the boundary
is locally smooth (a line segment, in fact), but at an uncountable set of
boundary points (of vanishing Lebesgue measure) the fractal dimension is
non-integral, and rather close to 2.

After these planar examples, let us briefly sketch the situation in 3D.
Clearly, there are the so-called T-phases (with T for ``thumbtack'', to
mimic their geometric structure) which \index{T-phase}
are quasiperiodic in a plane and periodic along the perpendicular line. 
They can be modelled by 3D tilings that are stacked layers,
each single layer being made from prisms (as tiles) whose base pattern forms 
one of the
classic planar tilings. Clearly, their Fourier image needs one extra Miller
index, i.e.\ the Bragg peaks of standard decagonal T-phases are indexed by
5 integers, 4 being needed for the non-periodic planar degrees of freedom
and one extra index for the periodic direction.

Of greatest importance probably are the tilings with icosahedral symmetry.
Here, one has to distinguish three different types. All can be obtained by
the projection method from hypercubic lattices in 6-space. There are three
different Bravais types of them, the primitive ($\ZZ^6$), the face-centred
($D_6^{}$), and the body-centred ($D_6^*$) one \cite{Schwarz}. 
The three different icosahedral classes, see \cite{Dun} for a detailed description, 
are then also called primitive (or $P$-type), face-centred ($F$-type)
and body-centred ($B$-type), respectively. Since no application of the $B$-type 
models are presently known, I'll skip details of them.
\index{icosahedral symmetry} 

The standard $P$-type tiling is made from two rhombohedra, an acute and an obtuse
one. It has, in various degrees of completeness, 
a long history \cite{Suck}, and was first
described by means of the projection technique in \cite{kramerneri}. 
It will be denoted by KN. The diffraction shows icosahedral symmetry and a 
clear scaling with inflation multiplier $\tau^3$. \index{Ammann-Kramer tiling}
This is characteristic of $P$-type structures, and makes
the distinction from $F$-type rather simple as the latter displays scaling with
an inflation multiplier\footnote{The meaning of this will become clear in the
Section on inflation symmetry.} $\tau$ (the same would be true of the $B$-type).

The more important class (in terms of applications) is that of $F$-type tilings.
One of the earliest examples is the zonohedral tiling by Socolar and Steinhardt 
\cite{SS}, abbreviated as SS. \index{Socolar-Steinhardt tiling}
It is built from four proto-tiles, namely the acute rhombohedron
met above, the rhombic dodecahedron, the rhombic icosahedron and the
famous triacontahedron, also known as Kepler's body \cite{GS}.
Another example was found by Danzer \cite{Danzer}, \index{Danzer tiling}
which is very closely related (i.e.\ locally equivalent) as we shall see later. 
Danzer's tiling (called DT) is built from 4 tetrahedra.
Finally, based on the projection technique, several other $F$-type tilings have 
been investigated, see \cite{BKPZ,KP} and references therein for details. 
The most important of those (called ${\cal T}^{*(2F)}$) is built from six 
tetrahedra and is again very closely related to the two tilings mentioned before
(SS and DT), although it contains more local
information -- a concept to be made more precise in a shortwhile.

Up to this point, no further mathematical details or concepts were needed
to get a first impression (see \cite{Peter} for a general construction
scheme). But for a better understanding of the structures,
their symmetries and some of the new features, we now have to dive a little 
deeper into the world of discrete geometry. In particular, we definitely need
some good tools to handle the zoo of possibilities. Later, we shall see that
the number of ``known'' examples, al least those with ``nice''
properties, is actually rather small, and can be handled
with little more difficulty than needed for the crystallographic patterns.

\section{LI-Classes and Symmetry}

One basic concept for the general analysis of global order
 properties of discrete structures is the equivalence concept 
 of {\em local indistinguishability\/},
 also known as local isomorphism\footnote{Since this term is occupied
 with a different meaning in other (connected) areas, I suggest to replace
 the word isomorphism by indistinguishability here, which also avoids
 the introduction of a new abbreviation.} \cite{levstein}.
Since the infinite (mathematical) structure is considered as an
 approximation to large but finite physical objects, it is
 natural to identify those structures which are locally
 indistinguishable on arbitrarily large but finite scales. 
Such structures are called locally indistinguishable, or locally isomorphic.
\index{local indistinguishability} \index{local isomorphism} \index{LI-class}

We will use this term frequently in the sequel, so for a precise
 definition we introduce some notation.
The mathematical objects we deal with are, most generally, discrete
 structures in Euclidean space, i.e., sets of (possibly decorated)
 bounded subsets of the space which are locally finite in the sense that
 each ball of finite radius meets only finitely many structure elements.
If $A$ is such a discrete structure, then we call an {\em $r$-patch} of
 $A$ each subset of $A$ which is completely contained in a ball
 of radius $r$.
Now, two structures, $A$ and $B$, are {\em locally indistinguishable}
 (or locally isomorphic) if
 each $r$-patch of $A$ is, up to a translation, also an $r$-patch
 of $B$ and {\em vice versa} (a moment's reflection reveals that,
 in the general case, this `vice versa' is necessary to get a proper 
 equivalence relation).
The corresponding equivalence class of a structure is called its
 {\em local indistinguishability class\/}, or LI-class, for short.

It should be emphasized that this formal definition does not quite
 reflect the intuitive description of the first paragraph, because
 we have insisted on identity of $r$-patches up to {\em translations\/}
 only, rather than up to more general Euclidean motions.
This more restrictive relation will prove useful for other concepts to be 
 introduced in the next Section, and especially for the definition
 of generalized point symmetries.

One of the most outstanding properties of the experimentally observed
 aperiodically ordered structures like quasicrystals is the occurence
 of crystallographically forbidden symmetries in their diffraction
 spectra, e.g., fivefold axes.
On the other hand, it is clear that, e.g., a 3D discrete structure can possess
 at most one axis of exact fivefold point symmetry in a given direction, 
 because otherwise
 there would be a dense set of such axes, which is impossible for a locally
 finite structure. \index{point symmetry} \index{generalized symmetry}
Therefore, to take into consideration also spatially homogeneous
 structures with non-crystallographic
 symmetry properties, one has to enlarge the symmetry concept slightly.
This is easily done with the help of LI as defined above: we say that
 the isometric linear transformation $T$ is a 
 {\em generalized point symmetry element\/} of the structure $A$ 
 if, and only if, $A$ and $T(A)$ belong to the same LI-class.

What do these concepts mean in the crystallographic case?
Firstly, one sees at once that a structure $A$ is locally indistinguishable from
 a crystallographic structure $B$ if and only if there is a translation
 vector $t$ which translates $A$ to $B$: $B=t+A$.
As a consequence, in the crystallographic case, the generalized point symmetry
 of a structure coincides with the conventional point symmetry, as
 it should do. 

{}For aperiodic patterns, the generalized symmetry is a proper extension
 of exact symmetry.
For example, the well-known Penrose tiling of the plane may have exact
 fivefold symmetry, i.e., there are precisely four
 representatives (up to translations)  in its LI-class
 which have one point of exact (global) fivefold symmetry each; 
 most of the members of the
 Penrose LI-class, abbreviated by LI(PT) from now on, have no exact symmetry at all.
The generalized point symmetry group of each elements of this LI class
 coincides, however, with the symmetry group of the regular decagon,
 i.e., it is the dihedral group\footnote{The symbol $D_n$ appears in two
 different meanings, once for the corresponding root lattice and once for the 
 dihedral group of order $2n$. Since both are standard in the literature, and
 misunderstandings unlikely, we stick to this convention.} $D_{10}$.

\section{Parametrization of LI-classes}

Having given the definition of an LI-class is not quite the same as understanding
its structure. The latter is, in fact, more complicated than one might expect. 
To see this, let us first consider the case of a crystallographic pattern $\cal P$: 
its LI-class consists of all its translates, and can thus be parametrized by 
the points of a fundamental domain of the corresponding lattice, $\Gamma$,
of translations (e.g., its Voronoi or Wigner-Seitz cell, to be specific)
because ${\cal P} = {\cal P} + t$ for all $t\in\Gamma$. In particular,
the LI-class $LI({\cal P})$ simply consists of one translation class of patterns.
\index{LI-class} \index{torus parametrization}

The correspondence between patterns in LI($\cal P$) and points of a fundamental
domain is called the {\em torus parametrization\/} of LI($\cal P$) because such a
domain, upon identifying $\Gamma$-equivalent boundary points, becomes
a {\em torus\/} of the dimension of the lattice. Clearly, the answer cannot be this
simple for non-crystallographic patterns. Here, LI($\cal P$) does not
only contain all translates of $\cal P$, but also all other patterns that can be 
obtained as limits of these (w.r.t.\ the obvious topology of patch-wise comparison) --
and ``most'' members are of the latter type.
In fact, for repetitive\footnote{The term repetitive means the following: for each
radius $r$, there is another radius, $R=R(r)$, such that each $\cal P$-patch of 
radius $r$ can be found in every $\cal P$-patch of radius $R$.} 
aperiodic patterns, the
LI-class contains uncountably many ($2^{\aleph_0}$) translation (even 
congruence) classes \cite{Sch} -- so, things are a lot more complicated, and a 
parametrization would be handy.

It is not known how to achieve this in general, but, for certain patterns, 
it is indeed possible. Among them are those point sets and tilings that
can be constructed by the projection method introduced in Section 3.  
Let us take another look at Figure \ref{topar}. It shows, in addition to the
ingredients needed to visualize the projection mechanism, a shaded square that 
represents
a fundamental domain of the lattice $\ZZ^2$. It becomes a torus on identifying
opposite faces in the usual way. Now, let us mark a special point of the cut space
$\EE$ by a handle (this can be thought of as a reference point for the pattern). 
If we move $\EE$ around, and with it the strip, we obtain
different cut and project sets for each position of the handle in the fundamental
domain, but we do not get anything new beyond it, due to the periodicity of the
embedding lattice. So, the points on the 2-torus parametrize different Fibonacci 
chains, and it is well known that one actually exhausts the entire LI-class this way
\cite{Sch}. This is the {\em torus parametrization\/} of quasiperiodic LI-classes
\cite{BHP}.

There is one subtlety which we have suppressed so far. The projection method is
unique as long as no lattice point lies on the boundary of the strip -- in which 
case the corresponding projected object is called {\em regular}
(or {\em generic}). Such regular members
form the majority of the LI-class. Situations where lattice points fall on the
boundary of the strip, in turn, correspond to the union of several {\em singular}
patterns, each of which can be seen as a limit of regular patterns. The
singular patterns also belong to the LI-class. 
In this new light, the torus parametrization is one-to-one
for regular members of the LI-class, but multiple-to-one for singular members.
This point will become important in any potential classification of LI-classes
beyond the torus parametrization. We will briefly come
back to this in the next Section. \index{regular patterns}
\index{singular patterns}

In our Fibonacci example, the parametrization of singular members is two-to-one.
An interesting question is how different from one another two such singular
chains are. The answer relies on the structure of the window.
As long as its boundary (the intersection of the strip with 
$\EE_{\rm int}$) is of vanishing Lebesgue measure, the different singular members
attached to the same torus parameter differ from one another only at places of
zero density -- hence, their identification is physically 
reasonable\footnote{For certain applications, it is advantageous to distinguish
regular and singular patterns and to adopt a topological point of view, e.g.\
for questions such as the spectra of Schr\"odinger operators, see \cite{Belli}
for details.}.
Vanishing Lebesgue measure of the boundary of the window, in turn, does not seem to
be too restrictive. In particular, the example of Figure~\ref{acc} is still included,
as are all other compact sets in $\RR^n$ with boundary of 
Hausdorff (or fractal) dimension $<n$.

The great advantage of the torus parametrization is its universality in the sense
that we can use the same torus for all projection structures attached to the same
embedding lattice and the same choice of the cut space $\EE$. 
Some individual properties of the LI-classes are then encoded
in the precise way the singular members behave, but they are usually less
important or even irrelevant for considerations such as symmetry, inflation invariance
etc. In particular, the torus parametrization allows us to find all members of an
LI-class showing {\em exact} invariance under a given symmetry operation, including
new types of symmetry such as inflation/deflation symmetry to be discussed later. 
This is based on lifting the symmetry operation under consideration to a mapping on 
the torus. Then, the number of fixed points can be determined by calculating certain 
determinants. For details, together with explicit examples and a full treatment 
of the physically relevant symmetries, see \cite{BHP,HRB}.

An extension of this analysis to groups of transformations (rather than single
operations) is possible, and it is instructive to look at subgroups of the
icosahedral group and their action on the three possible types of LI-classes,
see Table~\ref{icotab}. \index{symmetry analysis}
In each case, there are precisely 64 inversion symmetric
members of the LI-class, and they distribute in a very peculiar way on the
subgroups of $Y_h$, the full icosahedral group. To be more specific, there are
two rhombohedral tilings in LI(KN) with full $Y_h$ symmetry, one being regular
and one singular, while there are 4 such members in LI(DT), say, three regular
and one singular. This shows at least one reason why $F$-type structures are more
frequent than the other possibilities: as a consequence of this analysis, and using
the implications of local indistinguishability, it must be concluded that $F$-type
tilings or Delone sets have a denser distribution of clusters with exact (or almost
exact) icosahedral symmetry -- an idea pretty close to the concept of a
Frank-Kasper phase. This would suggest that $F$-type icosahedral quasicrystals
should be more frequent than $P$-type ones, as is indeed the case.

\begin{table}[ht]
\vspace*{1mm}
\centerline{\large
\begin{tabular}{|c|rrr|}   \hline
group    &  $B$-type & $P$-type & $F$-type \\
\hline
$Y_h$    &         1 &        2 &        4 \\
$D_{5d}$ &        18 &       12 &        0 \\
$D_{3d}$ &        30 &       20 &        0 \\
$D_{2h}$ &        15 &       30 &       60 \\
\hline
\end{tabular}  }
\vspace*{3mm}
\caption{Point symmetries of the 64 inversion symmetric tilings for the three
   different types of icosahedral LI-classes.} \label{icotab}
\end{table}

\section{Local derivability and MLD-classes}

Local indistinguishability (or local isomorphism), as defined above, is certainly
 a useful tool for the study of various properties of individual tilings.
However, for the description of more general aspects of their order
 (regardless of local details that might be more accidental), one has to
 extend the equiva\-lence concept under consideration \cite{BS}.
Let us make this vague statement a bit more concrete by
 considering a crystallographic structure.
Its order is, up to a scaling factor, fully described
 by its space group, without caring about how the fundamental domain
 is actually decorated. Also, it does not matter which representative
 of the fundamental domain we actually choose -- we can certainly transfer
 the detailed description from one choice to another in a strictly local way.
Now, in the aperiodic case, the (naively defined) space group is
 almost always trivial and therefore  {\em cannot\/}
  serve as a classification tool.
One way out is the consideration of the diffraction intensities of the
 structure, as has been done in \cite{Mermin,mermin2}.
Though this (non-rigorous!) approach works in a large class of structure types, 
 it has a number of shortcomings.
First, it is confined to the case of structures which show Bragg
 diffraction, i.e., are essentially almost periodic.
Next, the method is a little bit indirect, namely working in
 $k$-space, and offers therefore no good intuition for the things
 going on in real space.
Finally, such an approach might lead to a rather coarse picture, not
 distinguishing between locally inequivalent structures. This does not happen in
 the crystallographic case, but it does so in the extension of this
 approach to quasicrystals.

\begin{figure}[ht]
\vspace*{3mm}
\centerline{\epsfysize=35mm \epsfbox{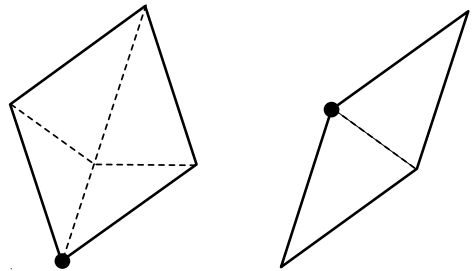}}
\vspace*{5mm}
\caption{Robinson's decomposition of the Penrose tiling.} \label{rdecom}
\end{figure}

To describe an appropriate alternative, let us consider, as an example, 
 two tilings which certainly are to be considered equivalent: the
 well known rhombic Penrose tiling (PT), and
 its Robinson decomposition (RD) into isosceles golden triangles,
 see \cite{BKSZ} and references therein for details.
By definition, there is a clear-cut rule to transform a PT into its corresponding
 RD, and it is not hard to see that, for a given RD, the underlying PT
 can be reconstructed, at each place, by observing just a few triangles
 in the vicinity, see Figures~\ref{pen}, \ref{rdecom} and \ref{trtil} (right half).
In particular, this means that we can transfer {\em any} decoration of RD into
one of PT, and vice versa -- eventually for the price that we have to distinguish
several congruent copies of the proto-tiles from one another, depending on their
(uniformly!) local neighbourhood in the tiling. 

Let us put this type of equivalence in more formal terms.
It is clear that the details of the transformation process described
 above are not important, e.g., we will certainly not care about
 tiles being properly dissected or composed, or in fact about tiles
 at all: quite frequently a representative discrete point set \cite{rich1} 
 is what one really needs or wants!
The essential feature which allows the abstraction from local details
 while keeping track of the global order is the {\em uniform locality\/}
 of the transformation rule.
It is easy to see that such a uniformly local rule for the transformation
 of some structure $A$ into a structure $B$ exists precisely under
 the following condition:
 There is a fixed finite radius $r$ such that if the $r$-patches
 of $A$ around two points, $p$, $q$, are equal up to the translation
 $t=p-q$, then the structure $B$ at the points $p$ and $q$ is the same, 
 again up to the translation $t$.
\index{local derivability} \index{local equivalence}

If this condition is fulfilled, then we call $B$
 {\em locally derivable} \cite{BSJ} from $A$.
If it is also fulfilled with the roles of $A$ and $B$ interchanged
 (with a possibly different radius $r'$) then we call $A$ and $B$
 {\em mutually locally derivable} from one another, or {\em locally equivalent}.

It is clear that this equivalence relation can be extended to entire
 LI-classes.
That is, if $A$ and $B$ are locally equivalent, then, for any $A'$
 in the LI-class of $A$, some $B'$ can be found in a canonical
 fashion (just using ``the same rule'') such that $A'$ and $B'$ are
 locally equivalent, thereby defining a one-to-one correspondence
 between the two LI-classes. \index{MLD-class}
Therefore, we can combine these equivalences defining the {\em MLD-class}
 of a structure $A$ to be the set of all structures which are 
 locally indistinguishable from some structure locally equivalent to $A$.
Needless to say that PT and RD in the example above belong to the
 same MLD-class in this sense. It is sometimes more useful to view MLD-classes,
which are defined as {\em unions} of LI-classes, directly as {\em sets} of 
LI-classes, but we will identify these two points of view for simplicity.
\index{Robinson tiling}

Let $A$ and $B$ be locally equivalent structures.
Obviously, if $A$ is invariant under a certain translation $t$, then
 $B$ must be invariant under $t$ as well, by the very definition of
 local equivalence.
A little further reflection shows that, on the other hand, if $A$ and
 $B$ are crystallographic with the same translation lattice, $\Gamma$, then 
 they are locally equivalent, by ``transformation rules'' involving
 only a couple of fundamental domains.
Therefore, local equivalence is a generalization of ``having the same
 translation lattice'' in the periodic case:
\begin{theorem}
   Two crystallographic patterns, $A$ and $B$, are locally equivalent
   if and only if they share the same translation lattice $\Gamma$.
\end{theorem}
This result also explains why we restricted our definition of local
indistinguishability (or patch-equivalence) to translations only, 
rather than using a version involving congruence. 
Below, we will refine the MLD concept in order
 to achieve a generalization of the space group classification.

\section{Local equivalence and limit translation module}

In general, it may be a hard problem to decide whether
 two given structures are locally equivalent or not.
As one tool for this task, we introduce an object associated to each
 discrete structure. It also generalizes the translation lattice of a periodic 
 structure in a certain sense, but contains less information\footnote{This
slightly more difficult Section may be skipped on first reading.}.

Let $A$ be a discrete structure.
For each radius $r$, one can collect all translation vectors $t$ which
 ``move patches inside $A$'', that is, for which an $r$-patch $P$ exists in
 $A$ such that the translate $t+P$ is also an $r$-patch of $A$.
These translation vectors for fixed $r$ generate a $\ZZ$-module,
 $\Lambda_r$, which simply consists of all integer linear combinations
 of the translations found. This module
 gets smaller if $r$ becomes larger: $r\leq r'$
 implies $\Lambda_r \supseteq \Lambda_{r'}$.
This property allows us to define a limit,
\begin{equation}
    \Lambda \; := \; \bigcap_{r>0} \Lambda_r
\end{equation}
which is again a $\ZZ$-module. We call $\Lambda$ the {\em limit translation module} 
(LTM) of the structure $A$. It is obviously an invariant of LI-classes.
\index{limit translation module}

In the crystallographic case, the LTM is just the translation lattice itself;
 on the other hand, in the general case, $\Lambda$ may turn out to be
 trivial, and this might even be the ``typical'' situation.
However, there is a large class of structures where this LTM is
 nontrivial and provides important information on the order of the structure,
 among them being point sets or tilings obtained by the projection method.
Anyway, the LTM is an invariant of MLD-classes: if $A$ and $B$ are
 locally equivalent, then their LTM must coincide; furthermore, if
 $B'$ is locally indistinguishable from $B$, then its LTM is the same as well.
On the other hand, if $B$ is derivable from $A$, one can only conclude
 that LTM($A$) $\subseteq$ LTM($B$).
Therefore, the determination of the limit translation module is a natural
 first step for proving or disproving the local equivalence of two
 given structures.

That the LTM does not specify the MLD class completely, even if it
 is non-degenerate, can also be studied in the special case of 
 cut and project patterns as introduced above.
If $A$ is such a quasi\-periodic structure, minimally embedded \cite{Sch}
 into the periodic structure $S$ (this is important!), then the limit 
 translation module
 of $A$ turns out to be just the projection image of the LTM of $S$
 (i.e.\ its lattice of translations, which then serves as the embedding
 lattice) into the subspace containing $A$, i.e.\ into $\EE$.
This also explains the connection of LTM($A$) to the Fourier module of $A$
 in such a situation: the latter is the (generally dense) set of points
 in $k$-space where we have to expect Bragg peaks. It is obtained by
 projecting the {\em dual} of the embedding lattice into $\EE$,
 as explained earlier.
But, given the higher-dimensional lattice and the cut space, the
 quasiperiodic structures obtained by cut-and-project are, by no means, all
 in the same MLD-class.

\begin{figure}[ht]
\vspace*{3mm}
\centerline{\epsfysize=50mm \epsfbox{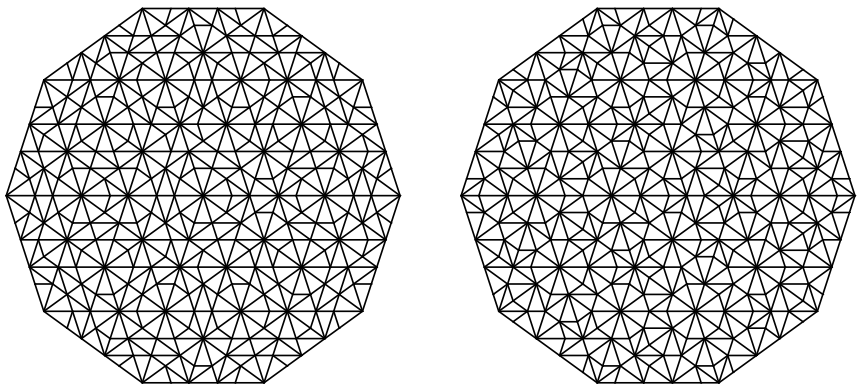}}
\vspace*{5mm}
\caption{Decagonal patch of the T\"ubingen triangle tiling TTT (left) versus
         Robinson's decomposition (RD) of the rhombic Penrose tiling (right).}
\label{trtil}
\end{figure}

In fact, one has the following simple necessary and sufficient criterion
 for two cut-and-project structures to share the same MLD-class: it must be
 possible to reconstruct the acceptance domains of $A$ from those
 of $A'$ by finitely many union, intersection and set complement operations,
 and vice versa (for a more detailed account of the relationship
 between the projection formalism and local derivability, see \cite{BSJ}).
This way, one can actually prove that PT and TTT (the so-called T\"ubingen
 triangle tiling, another decagonal tiling built from the golden triangles, 
 see Figure~\ref{trtil} and Ref.~\cite{BKSZ}) are {\em not} in the same
 MLD-class (although it is possible to rescale PT such that it becomes
 locally derivable from TTT). 
These two LI-classes actually differ in the distribution of singular tilings,
and the transformation rule from TTT to PT maps certain sets of singular
tilings of LI(TTT) onto single, but regular members of LI(PT) -- something that 
clearly cannot be inverted.
This is remarkable as TTT and PT certainly have
 the same space group according to \cite{Mermin}, i.e.\ they {\em cannot}
 be distinguished on the basis of the symmetry properties of their
 Fourier transforms. \index{T\"ubingen triangle tiling}

\section{Local derivability and symmetry preservation}

The careful reader will have noticed that we did not include rotations
 etc.\ in the definition of local derivability -- for very good reasons.
Nevertheless, symmetry is important, in particular for many physical 
 applications, and it is obvious how to include it by one further 
 step of refinement \cite{BS}.
Let us say that a certain local derivation rule preserves the symmetry
 of a structure if derivation rule and symmetry operation commute.
This then defines S-MLD-classes (for \underline{s}ymmetry preserving MLD)
 which obviously form (pairwise disjoint) subclasses of MLD-classes.

As in the case of MLD-classes, the specialization of the S-MLD concept
 to the periodic case fits well into classical crystallography:
Two crystallographic structures belong to the same S-MLD-class
 if and only if their space groups are identical (for a true generalization
 of the crystallographic classification scheme of periodic structures
 one has to broaden the S-MLD-classes by allowing for global
 similarity transformations).
Simultaneously, this fact shows that S-MLD is a proper refinement
 of MLD, i.e., there are MLD-classes which contain several S-MLD-classes.
An aperiodic planar example for such a behaviour is given at the end
 of the following Section. \index{MLD-class} \index{S-MLD-class} 

In 3-space, Danzer's tiling (DT) and that of Socolar and Steinhardt (SS) are in the
same S-MLD-class, see \cite{Roth,DPT} for a proof. So, in this sense, they
really describe the same class of structures. One of the $D_6$ based tilings
allows the local derivation of Danzer's tiling from it, but there is no local
rule to go back -- a 3D analogue of the situation previously met with
TTT versus PT.

\section{Inflation symmetries and matching rules}

Let us now investigate how the MLD-concept works in the context 
 of two outstanding properties which important aperiodic structures exhibit: 
 {\em inflation/deflation symmetry\/} and {\em perfect matching rules\/}
 (compare \cite{katzmatch,ltqt,levi,petermatch,KG} for commonly used definitions).

\begin{figure}[ht]
\vspace*{3mm}
\centerline{\epsfysize=35mm \epsfbox{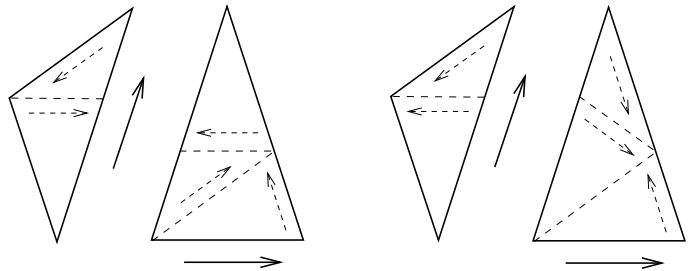}}
\vspace*{5mm}
\caption{Inflation rule for TTT (left) and RD (right).} \label{defttt}
\end{figure}

As there are various concepts of inflation/deflation in the literature, we
 have to make precise what we mean by it, thereby taking the opportunity to put 
 the MLD concept into operation.
Usually, an inflation of a structure consists in a certain rule for a local
 transformation of structure elements into patches of a new structure which
 turns out to be of the same type as the original one, but on a smaller scale.
For example, the dissection of the golden triangles depicted in Figure~\ref{defttt}
 gives the inflation rules both for the triangles of TTT and of RD.
So far, this does not seem to be too interesting, as one can do this
 sort of procedure with the periodic tiling of the plane by squares.
However, in certain cases, such an operation does not result in {\em any} loss
 of information on the original structure, i.e., it is possible to recover
 it by an inverse transformation, also in a local fashion.

Rereading the definitions in the last Sections, one sees that the above
 description is precisely what is meant by the following formal definition:
 a structure $A$ has an {\em inflation/deflation symmetry\/} related
 to a similarity transformation $T$, if $T(A)$ is in the same MLD-class as $A$
 (i.e., if $T(A)$ is locally indistinguishable from a structure which, in turn, is
 locally equivalent to $A$; it is necessary to phrase it this way, because the
 situation where already $T(A)$ is locally equivalent to $A$ is too special).
\index{inflation/deflation symmetry} 

From the last Section, we may conclude that no
 periodic structure can have {\em any} in-/deflation symmetry related
 to a nontrivial $T$, i.e., where $T$ is not just a rigid motion.
The reason can be glimpsed from the following example: subdividing the
 square cells of the lattice $\ZZ^2$ into smaller squares of half the edge
 length, say, is obviously a local rule. But the converse, re-grouping 
 four adjacent squares into a bigger square, is not -- it requires the
 knowledge where the process was started to guarantee fault-free operation,
 and this means it is not possible by a local rule.
On the other hand, the existence of nontrivial in-/deflation seems to
 be a very common feature among the interesting aperiodic
 structures \cite{petermatch}.
In many cases, $T$ is just a rescaling, but there are important
 examples where rotation-dilations are needed, as in certain 2D tilings with
 12-fold symmetry (cf.\ Refs.\ \cite{niiz,BJS}). We have
\begin{theorem}
 The existence of an inflation/deflation symmetry is a property of an 
 entire MLD-class: either all members share it, or none has it.
\end{theorem}
This fact may serve into two directions.
Firstly, having established the existence of in-/deflation for a single structure,
 one already has in-/deflation for its entire MLD-class.
Secondly, simultaneous (non-)existence of in-/deflation provides
 a necessary criterion for two structures to be locally equivalent.

The study of perfect matching rules is another subject where the concept
 of local equivalence proves fruitful.
We say that a structure $A$ possesses {\em perfect matching rules\/} (essentially
 in the sense of Ref.\ \cite{levi}) if its LI-class is determined by the
 set of its $r$-patches for some {\em finite\/} radius $r$ (which we call, if
 chosen minimally, the matching rule radius of $A$), i.e., if every
 other structure which contains, up to translations, only $r$-patches
 which also occur in $A$ necessarily belongs to the LI-class of $A$.
In the case of LI(PT), a very simple version in terms of tiles with oriented
 edges can be given, see \cite{Uwe} for an illustration. 
For obvious reasons, this property of a structure is particularly
 interesting in the case that this structure is supposed to
 describe the global order of physically realized structures as
 quasicrystals \cite{katzmatch,KG}. One may think of a Hamiltonian that favours
 the patches of the atlas, this way restricting the groundstate to a member
  of the LI-class. \index{matching rules} \index{perfect matching rules}

It is almost immediate that every structure which is locally
 equivalent to one with perfect matching rules must itself possess
 perfect matching rules:
\begin{theorem}
  The existence of perfect matching rules is a property of an 
  entire MLD-class: either all LI-classes contained in this MLD-class 
  possess perfect matching rules, or none of the LI-classes can have them.
\end{theorem}
Note, however, that the matching rule radius is {\em not} an invariant of 
 MLD-classes. This is of some relevance in the physical context, if one
 tries to relate the matching rules to the local interaction of some suitable
 Hamiltonian, i.e.\ if one searches for a Hamiltonian whose ground states
 form a specific LI-class with perfect matching rules.
This is due to the fact that the information contained in a structure
 may be delocalized (gradually) by the local derivation of another structure.
An estimate for the matching rule radius of a structure $A$ which is
 locally equivalent to a structure $B$ with perfect matching rules
 is given in \cite{fmm}. It involves the matching rule radius of $B$ and the 
 relevant radii for the transition from $B$ to $A$ and vice versa.
It is an interesting question what the infimum of all matching rule radii
of the LI-classes inside one MLD-class is. It has been conjectured that
it might actually be zero under some extra condition. This is rather plausible
for systems with inflation-deflation symmetry, as shrunk-down representatives
exist on arbitrarily small scales.

It should be noted that there are certain tilings, such as the Ammann-Beenker
 octagonal tiling \cite{Ammann} or G\"ahler's dodecagonal ``shield''
 tiling \cite{franz}, which do {\em not} possess perfect matching rules
 if one considers only undecorated tiles, but can be transformed
 into structures with matching rules by convenient decorations.
In these cases, the introduction of the decoration cannot
 be achieved in a local fashion (see Ref.\ \cite{franz}),
 the naked and decorated
 tilings form different MLD-classes and should therefore be
 distinguished clearly \cite{rich2}.

One might ask for the number of different possibilities to
 construct tilings of a given symmetry with perfect matching rules
 (for a study of the 8-, 10-, and 12-fold symmetrical cases
 see Ref.~\cite{lueckmatch}).
Here, one is not interested in the (infinite) variety
 of representatives of one and the same MLD-class, but in {\em different}
 MLD-classes with perfect matching rules, such as those defined by
 LI(PT) and LI(TTT). 
At the moment, the only candidate of an infinite family of different
 tilings with perfect matching rules is provided by the generalized
 Penrose patterns with parameter $\gamma=m+n\tau$, see \cite{ltqt} for details. 
A closer inspection, compare also Ref.~\cite{inger,ltqt}, shows
 that these tilings belong to only one MLD-class,
 or to two S-MLD-classes -- one with fivefold and one with tenfold
 symmetry (which contains LI(PT)). This analysis has been extended in
\cite{ltqt} to rational values of $\gamma$, which results in an entire tower
of LI-classes that allow a local derivation down the tower, but not upwards.
On the bottom of this tower, we find a well-known friend: the rhombic
Penrose tiling, PT. Similar towers certainly exist for other examples,
e.g.\ in 3-space, but, to our knowledge, have not yet been analyzed in detail.
\index{generalized Penrose tilings}

\section{Summary of the perfect world}

So far, it has been outlined, in rather elementary terms,
 how equivalence concepts such as that of local indistinguishability (LI)
 and that of mutual local derivability (MLD)
 are helpful in sorting out local properties of locally finite
 tilings or other discrete patterns.
Furthermore, we believe that concepts along the lines presented
 above are needed to continue a sound classification program
 of aperiodically ordered structures, and for various aspects it is advantageous
 not to depend on Fourier transforms.

Let us continue with a speculation. We have seen that there were serious
connections between different examples of patterns with perfect matching rules.
An interesting question is how serious these connections are. 
From the past fifteen years of research on quasicrystals and aperiodic
order, we have the feeling that the variety of S-MLD-classes
 with {\em all} magic properties is limited, if organized properly.
To this end, one has to form towers of them, in which two consecutive members
allow a local derivation down the tower, but not up -- as in the case of
MLD(TTT) being on top of MLD(PT). With this,
we tend to the following conjecture:
 the number of towers of quasiperiodic S-MLD-classes with fixed symmetry, limit
 translation module of minimal rank\footnote{Minimal w.r.t.\ the symmetry, e.g.\
 rank 4 for tenfold symmetry in the plane or rank 6 for icosahedral symmetry
 in 3-space.}, local inflation/deflation symmetry (with fixed inflation multiplier) 
 and local perfect matching rules is finite.
\index{classification scheme} \index{tiling towers}

This statement is rather fragile: removing essentially any of its conditions,
it is wrong. So, a further exploration of this question (and a proof or disproof)
 would be a logical next step in the classification of
 aperiodic structures -- complementing and perhaps even completing
 the existing classification of Fourier modules 
 \cite{Ted,mermin2} (mainly based on symmetry alone).

Although a good classification of order, even along these lines, is not in
sight (and it might actually be very far away), one should not close ones eyes
in front of other possibilities. In fact, even if we had the answer to the
above question, it would not be sufficient physically: so far, we have totally,
and deliberately, ignored any {\em stochastic} aspect of point sets or tilings.
This is not really tolerable, and the last Section is now devoted to a very
brief and sketchy introduction to a totally different (though, fortunately,
not totally disconnected) universe ...

\section{Alternatives: disorder and random tilings}

The first indication on incompleteness of the above approach comes from the observation
that also quasicrystals will show defects (in fact, probably more than ordinary
crystals), and one would like to know the possible scenarios. As a first step, one
can investigate so-called defective vertex configurations within the geometric
setting of a tiling, compare \cite{Shelomo} and references therein for a survey.
Although this provides rather interesting insight, and even allows for some
simplistic, but not too unrealistic, models, we shall focus here on an alternative
to perfect tilings or Delone sets that starts from a rather different, if not
antipodal, point of view. \index{defects}

The above discussion implicitly concentrated on discrete structures that could be seen
as idealizations of {\em energetically\/} stabilized structures, 
and LI-classes with perfect
matching rules may be interpreted in this direction: in principle, there exists a
Hamiltonian whose ground state is necessarily an element of the LI-class defined
by the perfect matching rules. Although this idea seems attractive, it has a number
of pitfalls. On the one hand, such a Hamiltonian will be rather artificial and
unrealistic, and, on the other hand, such matching rules do not lead to
local growth rules \cite{Uwe,Gerrit} -- unless the structure is crystallographic.
So, even if matching rules can provide a toy model for energetic stabilization,
they cannot explain why quasicrystals form and how they grow. A similar problem
is also present in a slightly relaxed scenario, known as ``maxing rules'', see
\cite{Petra,JS}. \index{entropy} \index{stabilization}

An alternative, or complementary, idea is provided by statistical physics.
Recall that stability is related to minimizing the free energy which is given
as $F = U - T S$, with $U$ the internal energy, $S$ the entropy and $T$ the
temperature. Therefore, stabilization can also have a significant
{\em entropic} contribution, and we will now briefly sketch some ideas that have been
developed along that route. To this end, we assume that the internal energy, $U$,
is essentially degenerate, and minimizing $F$ thus means maximizing $S$. So, we are
talking about an idealization (first pointed out by Elser \cite{Veit})
which corresponds to a high-temperature phase -- a
picture that is not at all absurd in view of the experimental observations!
\index{random tilings}

Let us go back to the Fibonacci chain for a moment. It showed two different atomic
distances between neighbouring points, cells $a$ and $b$, say. 
Their arrangement is entirely deterministic, with frequencies $1/\tau$ and
$1/\tau^2$, respectively. Furthermore,
one can show that precisely $n+1$ different words (in $a$ and $b$)
of length $n$ occur in an infinite Fibonacci chain -- no matter which representative
of its LI-class we take. So, its (combinatorial) entropy density per letter 
vanishes, as can easily be calculated:
\begin{equation}
      s \; = \; \lim_{n\rightarrow\infty} \frac{\log(n+1)}{n}
        \; = \; 0 \, .
\end{equation}
The same phenomenon happens for the entropy density 
of all other deterministically ordered structures obtained by the
projection method or by an inflation rule \cite{Berthe}. This is caused by the
inherent long-range repetitive order of those structures, which is also the 
reason for a well-defined diffraction spectrum\footnote{Inflation generated
tilings need not have pure point diffraction spectrum. This depends
on the nature of the inflation multiplier, see \cite{Boris} for a detailed
account.}. So, if we want to introduce some entropy, we have to go
beyond such structures, but we certainly want a scheme
that does not totally destroy the long-range (orientational) order.

The simplest idea is to relax the possible configurations of cells.
If we allow any sequence of $a$'s and $b$'s subject to the sole restriction that
their frequencies $\nu^{}_a$ and $\nu^{}_b$ match those of the Fibonacci chain, 
we obtain a Bernoulli-type ensemble of infinite chains, this time with positive 
entropy density
\begin{equation}
    s \; = \;  -\nu^{}_a \log(\nu^{}_a) - \nu^{}_b \log(\nu^{}_b)
      \; = \; \frac{\tau+2}{\tau+1} \log(\tau) \; \simeq \; 0.665 \, .
\end{equation}
This (essentially unrestricted) ensemble has two disadvantages: first, the
maximal entropy would occur for $\nu^{}_a=\nu^{}_b=1/2$ as $\log(2)\simeq 0.693$,
and not for the Fibonacci frequencies. Second, the new ensemble is so random that
the typical ensemble member does no longer show sharp diffraction peaks.
Consequently, this model is not suitable to explain Fibonacci-type structures.

To overcome these difficulties, at least in dimensions two and higher,
the idea of a {\em random tiling\/} was put forward by the Cornell group, 
see \cite{Henley} and references therein. Starting from the same
proto-tiles as in a perfect quasiperiodic tiling, one allows all gap-less 
space-fillings that are face-to-face and overlap-free, and eventually subject to
further restrictions. It is this set of local, geometric constraints 
which transforms the unrestricted Bernoulli-type ensemble into a more
interesting and realistic Markov-type ensemble.
E.g., one could start from the two rhombi of the Penrose
tiling of the plane but relax the matching rules. This way, one obtains a random
tiling ensemble with unique entropy maximum at the tile frequencies of the perfect 
tiling, and this maximum also corresponds to the unique point (in parameter space)
of maximum symmetry, $D_{10}$. 
So far, the situation seems considerably improved.
However, the diffraction side is not yet fully satisfactory: such planar
tiling ensembles display sharp peaks, but they are, in general, no Bragg peaks.

\begin{figure}[ht]
\vspace*{3mm}
\centerline{\epsfysize=4.2cm \epsfbox{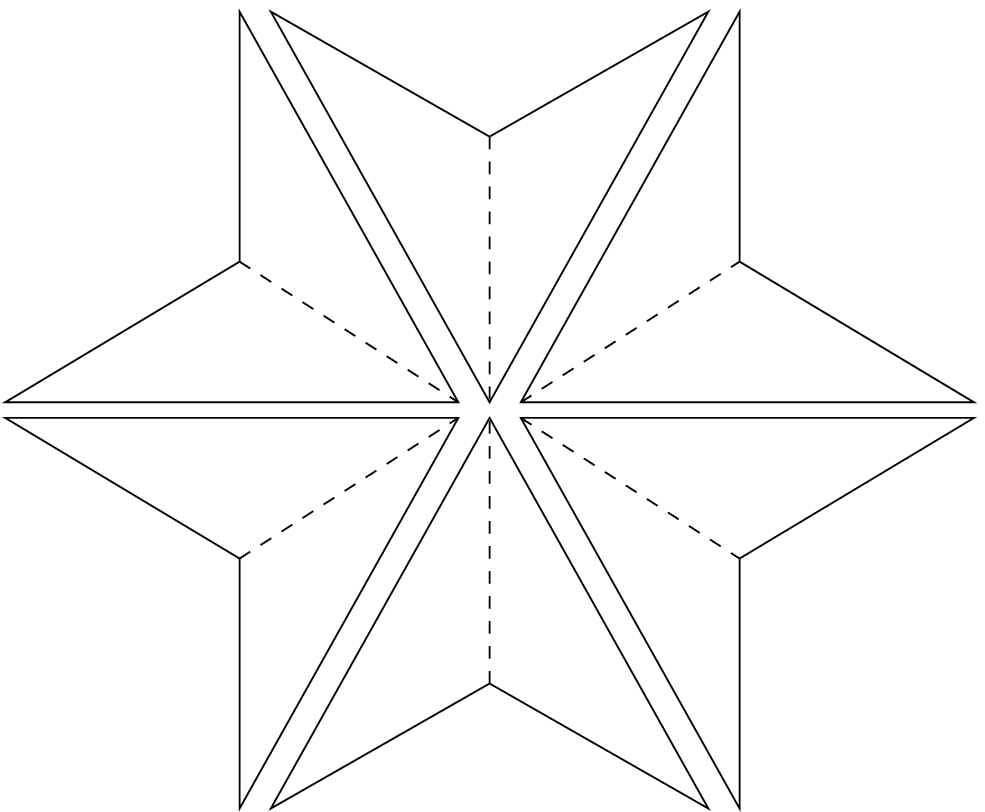} \hspace{2cm}
            \epsfysize=3cm \epsfbox{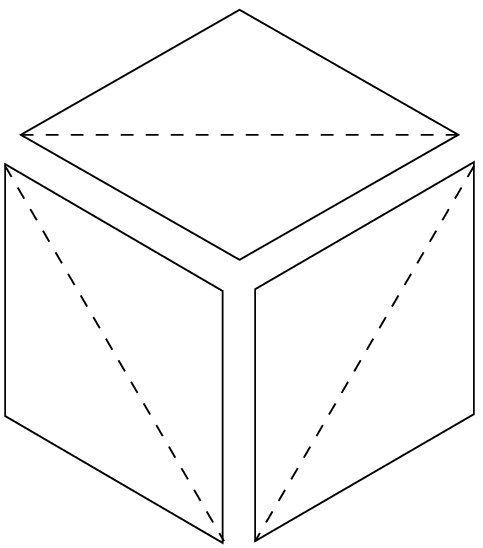}}
\vspace*{5mm}
\caption{Prototiles of the dart-rhombus random tiling.} \label{proto}
\end{figure}

Since we live in three dimensions, we should ultimately be concerned only with that
situation. But then, doing the analogous exercise with the tilings of 3-space by the
two rhombohedra mentioned earlier, we find again the entropy maximum at the point in
parameter space that shows icosahedral symmetry. This time, because statistical
fluctuations of a 3D ensemble are bounded, the diffraction would still show Bragg
reflexes, plus a structured diffuse background. Furthermore, the entropy around
the maximum, as a function of the parameters (e.g., the densities of different
proto-tiles), is a strictly
concave function with the maximum being locally quadratic. This allows to develop
a kind of entropic elasticity theory around the maximum, and the corresponding
symmetry adapted invariant forms are characterized by what are called the
{\em elastic constants} \cite{Henley}. \index{elastic constants}

We have met the basic ingredients. The collection of all tilings of $n$-space from a 
given set of proto-tiles (face-to-face, gap-less and overlap-free, as usual,
plus eventually some extra constraints) is
called a {\em random tiling ensemble\/} if it has positive entropy density in the
sense that the number of possibilities to cover a ball of radius $r$ grows
exponentially in $r^n$. Such ensembles, if well-defined statistically, have
some rather general features, two of which are often stated as kind of
axioms. They are the so-called random tiling hypotheses \cite{Henley} which, in 
our language, read \index{random tiling hypotheses}
\begin{description}
\item[RTH1] The point of maximum entropy is automatically a point of maximum symmetry.
\item[RTH2] The entropy, around the maximum, is a locally quadratic function of
            the parameters.
\end{description}
It might be interesting to note that RTH1 is actually a theorem and follows by
some group theoretic arguments from considerably simpler assumptions. 
The status of RTH2, however, is more difficult: this really is an assumption that
has to be verified, as explicit counterexamples are known, see \cite{RHHB}
for details.

\begin{figure}[ht]
\vspace*{1mm}
\centerline{\epsfysize=100mm \epsfbox{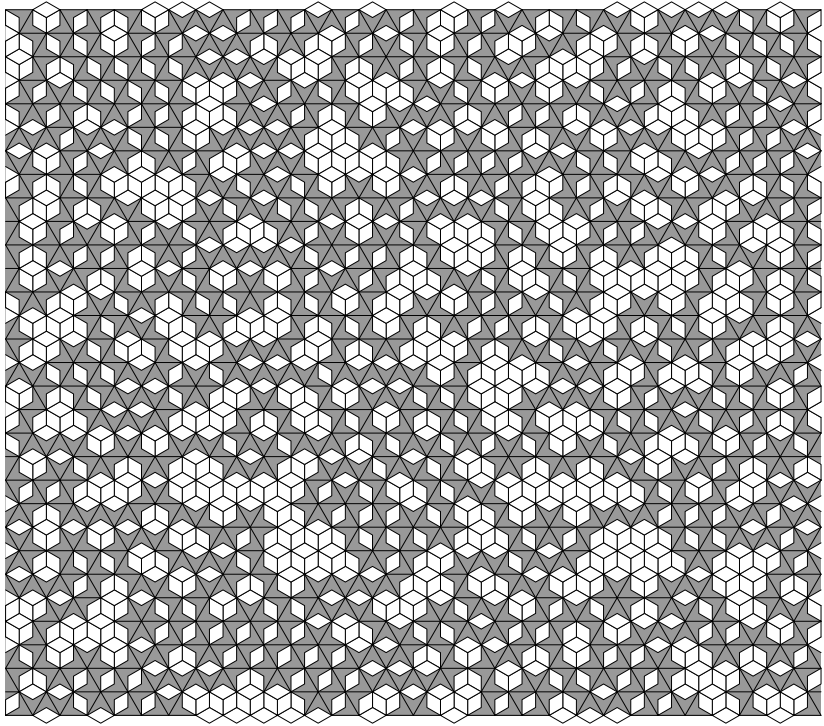} \hspace*{5mm}}
\vspace*{5mm}
\caption{Snapshot of a dart-rhombus random tiling at maximum entropy.}
\label{rantil}
\end{figure}

To give an impression, let us, at this point, present one example of a planar
random tiling, made from two proto-tiles, a rhombus and a dart, see Figure \ref{proto}.
Both proto-tiles are built from two copies of an isosceles triangle, and thus share
the same area. Rhombi occur in 3 orientations, darts in 6.
In addition to the usual rules to put them together, we also demand
that the rhombi observe an alternation rule, i.e., no two rhombi that are mere
translates of one another are allowed to share an edge. 
Also, two neighbouring darts must not share a short edge.
This is then one of the very
few non-trivial examples that can be solved completely, in an exact and even
rigorous way, by methods from statistical mechanics. Although it is only compatible
with sixfold symmetry (and could thus be called a crystallographic random tiling
ensemble), it shows many features that are also typical of 
``quasicrystalline'' random tilings (though the diffraction is different). 
In particular, it displays
a unique entropy maximum of quadratic nature [RTH2] at the point of maximal 
symmetry [RTH1]. \index{dart-rhombus random tiling}
Furthermore, somewhere else in the phase diagram,
it shows an order-disorder phase transition (of Onsager type)  
which can alternatively be interpreted as a percolation
phase transition, compare \cite{Hoeffe,RHHB} for details. In Figure \ref{rantil} we
show a typical snapshot from the ensemble at its entropy maximum, which turns
out to be 
\begin{equation}
    s \; = \; {1\over 3} \log(2) \; \simeq \; 0.231 \, .
\end{equation}
This result is so simple that it must be possible to give a direct derivation
of it, without referring to the somewhat elaborate exact solution. This is indeed
possible, see \cite{RHHB} or chapter 3.4 of \cite{Hoeffe}.

\noindent

\section{Concluding remarks}

After this little excursion into the world of aperiodic order, one should
realize that the displayed material was not only a short (and possibly
insufficient) glimpse, it was also highly biased. This resulted not so much
in a strange view at the things discussed, I hope, but in the omission of
many other aspects, often fascinating, sometimes frustratingly complicated,
occasionally in the t2c2e-category\footnote{Things too complicated to explain,
at least on a few pages.}, or just unnoticed by the ignorant author.
But, even fifteen years after the field took off, it is still in its infancy, and
any other statement bares the risk of delusive security. It is amazing
how many things seem to be known, or almost known, and how few of them
are really established, or fully known. And for many aspects, we have
just scratched the surface.

In view of this, I should try to mention some of the points not discussed above.
However, even that would be biased again, and it is probably better to give the
advice to read on in this volume, and to consult the introductory articles from
other summer or winter schools, some of which are given in the references below.
A more complete list of books and proceedings can be found in a short extra
chapter of this book, together with some guiding comments on what one can find
in the titles listed. Anyway, one ``take-home'' message should be to stay
open-minded to new aspects, to challenge ``common knowledge'', to follow
interesting paths, and -- last not least -- to have fun!

\section*{Acknowledgements}

This survey is partially based upon results obtained in collaboration with a number 
of colleagues, in particular with Uwe Grimm, Dieter Joseph, Peter Kramer, 
Robert V.\ Moody, Peter A.~B.\ Pleasants and Martin Schlottmann. 
Many results would not be in their present form without this cooperation. 
Also, I am grateful to my students Moritz H\"offe, Joachim Hermisson and 
Christoph Richard for their help and supply of material. Finally, I'd like to thank
Shelomo I.\ Ben-Abraham, Veit Elser, Franz G\"ahler, Albertus Hof, 
Jeffrey C.\ Lagarias and Hans-Ude Nissen for valuable discussions.
Many other names should be mentioned here, but I apologize for not doing
so and hope that the selection of references is a valid substitute for it.

\end{document}